\documentstyle[preprint,aps,eqsecnum]{revtex}

\def\beq#1{\begin{equation} \label{#1}}
\def\eeq{\end{equation}}
\newcommand{\bea}{\begin{eqnarray}}
\newcommand{\eea}{\end{eqnarray}}
\def\bra#1{\left\langle #1\right\vert}
\def\ket#1{\left\vert #1\right\rangle}
\def\epsp{\epsilon^{\prime}}
\def\NPB{{ Nucl. Phys.} B}
\def\PLB{{ Phys. Lett.} B}
\def\PRL{ Phys. Rev. Lett.}
\def\PRD{{ Phys. Rev.} D}

\begin{document}
{
\tighten

\title {Theory of neutrino oscillations using condensed matter physics\\
 Including production process 
and energy-time uncertainty}
\author{Harry J. Lipkin\,\thanks{Supported in part by
U.S.
Department of Energy, Office of Nuclear Physics, under contract
number
DE-AC02-06CH11357.}}
\address{ \vbox{\vskip 0.truecm}
  Department of Particle Physics
  Weizmann Institute of Science, Rehovot 76100, Israel \\
\vbox{\vskip 0.truecm}
School of Physics and Astronomy,
Raymond and Beverly Sackler Faculty of Exact Sciences,
Tel Aviv University, Tel Aviv, Israel  \\
\vbox{\vskip 0.truecm}
Physics Division, Argonne National Laboratory
Argonne, IL 60439-4815, USA\\
~\\harry.lipkin@weizmann.ac.il
\\~\\
}

\maketitle

\begin{abstract}

Neutrino scillations cannot arise from an initial  isolated one particle state
if  four-momentum is conserved. The  transition matrix element is generally
squared and summed over all final states with no interference between
orthogonal final states. Lorentz covariant  descriptions  based  on
relativistic quantum  field theory   cannot describe interference between
orthogonal  states with different $\nu$  masses producing neutrino
oscillations. Simplified model presents rigorous derivation of  
handwaving argument about ``energy-time uncertainty". Standard time-dependent
perturbation theory for decays shows how energy spectrum of final state
is much broader than natural line width at times much shorter than 
decay lifetime. Initial state containing two components with different
energies decay into two orthogonal states with different $\nu$ masses 
completely separated  at long times with no interference. At short
times the broadened energy spectra of the two amplitudes overlap and interfere. 
``Darmstadt oscillation" experiment attempts to
measure the momentum difference between the two contributing coherent 
initial states and obtain information about $\nu$ masses without
detecting the $\nu$. Simple interpretation  gives value for the
squared $\nu$ mass difference differing by less than a factor of three from
values calculated from the KAMLAND experiment. Treatment  
holds only in laboratory frame 
with values of energy, time and momentum determined
by experimental environment at rest in the laboratory.

\end{abstract} } 

\def\beq#1{\begin{equation} \label{#1}}
\def\eeq{\end{equation}}
\def\bra#1{\left\langle #1\right\vert}
\def\ket#1{\left\vert #1\right\rangle}
 \def\epsp{\epsilon^{\prime}}
\def\NPB{{ Nucl. Phys.} B}
\def\PLB{{ Phys. Lett.} B}
\def\PRL{ Phys. Rev. Lett.}
\def\PRD{{ Phys. Rev.} D}

\section {Introduction -
The basic paradox of neutrino oscillations}
\subsection
{\bf The problem}

\begin{enumerate}

\item {The original neutrino experiment by Lederman et al \cite{leder} showed a
neutrino emitted in a  $\pi \rightarrow \mu \nu$ decay entering a detector and
producing only muons and no electrons.}

\item {The neutrino enters detector as coherent
mixture of mass eigenstates with right relative
magnitudes and phases to cancel the amplitude for producing electron at
detector.}



\item{$\nu$ wave function must have
 states with  different masses, momenta and/or energies  
.}

\item{In initial one-particle state components with different momenta
have different energies.}

\item{Lederman et al experiment can't exist if energy and momentum are
conserved}

\end{enumerate}
\subsection
{\bf The Solution}
\begin{enumerate}
\item{If momentum is conserved in the interaction,
violation of energy conservation needed.}

\item{Energy-time uncertainty in the laboratory frame allows components of
initial wave packet with different energies to
produce same final $\nu_e$ with the same single energy.} 
\end{enumerate}

\subsection
{\bf Darmstadt application} 

Radioactive ion circulates in storage ring before decay\cite{gsi}

\begin{enumerate} 
\item {Transition probability depends on relative phase between two components}

\item {Relative phase and transition probability change in propagation through
storage ring.}

\item {Phase changes produce oscillations in decay probablity}

\item{Oscillations can give information about $\nu$ masses without
detecting the $\nu$ .}

\end{enumerate}

\subsection {A simple example of resolution of the paradox}

Time dependent perturbation theory shows violation of energy conservation                                                     
by energy-time uncertainty in sufficiently short times\cite{qm}.  
The time dependent amplitude  $\beta_f(E_i)$  
for the decay from an initial state with energy $E_i$ 
into a final state with a slightly different energy $E_f$ is           
\beq{timepert}
\frac{\beta_f(E_i)}{g}\cdot(E_i - E_f)=
\left[e^{-i(E_i-E_f)t}-1]\right]\cdot e^{-2iE_ft}
\eeq
where
we have set $\hbar=1$ and $g$ is the interaction coupling constant.

We now generalize this expression to the case where two initial states with
energies $E_f - \delta$ and $E_f + \delta$ decay into the same final state with
energy $E_f$    
and define $x\equiv E_i-E_f$
    
\beq{timepert2b}
\frac{e^{2iE_ft}}{g}\cdot [\beta_f(E_f  + x - \delta) +  \beta_f(E_f + x+\delta)]=
\left[\frac{e^{-i(x - \delta )t}-1}{(x - \delta)}\right] +
\left[\frac{e^{-i(x + \delta )t}-1}{(x + \delta )}\right] 
\eeq

The square of the transition amplitude denoted by $T$ is then given by

\beq{timepert5 }
\frac{|T^2|}{g^2}\equiv \left[\frac{\beta_f(E_f + x - \delta) +  
\beta_f(E_f +x + \delta)}{g}\right]^2 =
4 \cdot \left[\frac{\sin^2 [(x- \delta)t/2]}{(x-\delta)^2} + 
\frac{\sin^2 [(x+ \delta)t/2]}{(x + \delta)^2}  \right] + T_{int}
\eeq
where the interference term $T_{int}$ is

\beq{tint0}
T_{int} = 
\left[\frac{e^{-i(x - \delta )t}-1}{(x - \delta)}\right] \cdot
\left[\frac{e^{i(x + \delta )t}-1}{(x + \delta )}\right]  + cc =
4\left[ \frac{2\sin^2 [\delta t/2] + 2\sin^2 [x t/2]\cos [\delta t]
- \sin^2 (\delta t)}
{x^2 - \delta^2}  \right]
\eeq   

If the time is sufficiently short so that the
degree of energy violation denoted by $x$ is much larger than the energy
difference $\delta$ between the two initial states, $x \gg \delta$ and 

\beq{tint2}
x \gg \delta; ~ ~ ~
|T^2|\approx
8 g^2 \cdot \left[\frac{\sin^2 [xt/2]}{x^2}\right]\cdot[1+\cos \delta t]  
\eeq

The transition probability is given by the Fermi Golden Rule. 
We integrate the the square of the transition amplitude over $E_i$ or
$x$, introduce the density of final states $\rho(E_f )$ and                     and
assume that $\delta$ is neglibly small in the integrals.

\beq{timepert6 }
\int_{-\infty}^{+\infty} |T^2| \rho(E_f ) dx \approx
\int_{-\infty}^{+\infty}
 8  g^2 \cdot \left[\frac{\sin^2 [xt/2]}{x^2}\right]\cdot[1+\cos \delta t] 
   \rho(E_f ) dx 
\eeq
The transition probability per unit time $W$ is then
\beq{tranprob}
   W  \approx 4  
g^2 \cdot \int_{-\infty}^{+\infty} du \left[\frac{\sin^2 u}{u^2}\right]\cdot
  \rho(E_f )
[1 +  \cos (\delta t)]\cdot t = 4 \pi g^2 \rho(E_f )
 \eeq

The interference term between the two initial states is seen to be comparable to
the direct terms when $\cos (\delta t) \approx 1$; i.e. when the energy
uncertainty is larger than the energy difference between the two initial states.

This example shows in principle how two initial states with a given momentum
difference can produce a coherent final state containing two neutrinos with the
same energy and the given momentum difference. A measurement of the momentum 
difference between the two initial states can provide
information on neutrino masses without detecting the neutrino.

In this simple example the amplitudes and the coupling constant $g$ are assumed
to be real. In a more realistic case there is an additional extra relative phase
between the two terms in eq.(\ref{timepert2b}) which depends upon the initial
state wave function. In the GSI experiment\cite{gsi} this phase varies linearly
with the time of motion of the initial ion through the storage ring. This phase
variation can produce the observed oscillations.

\section {The basic physics of neutrino oscillations}

\subsection {Interference is possible only if we can't know everything}

Neutrino oscillations are produced from a coherent mixture of different  $\nu$
mass eigenstates. The mass of a $\nu$ produced in a reaction where all other
particles have definite momentum and energy is determined by
conservation of energy and momentum. Interference between amplitudes from
different $\nu$ mass eigenstates is not observable in such a ``missing mass"
experiment. Something must prevent knowing the neutrino mass from conservation
laws. Ignorance alone does not produce interference. Quantum mechanics must
hide information. To check how coherence and oscillations can occur we
investigate what is known and what information is hidden by quantum mechanics.

A simple example is seen in the decay $\pi \rightarrow \mu
+ \nu$. If the momenta of the initial $\pi$ and recoil $\mu$ are known the
$\nu$ mass is known from energy and momentum conservation and there
are no oscillations. But oscillations have been observed in
macroscopic neutrino detectors at rest in the laboratory system. 
Oscillations arise
only  when the outgoing $\nu$ is a wave packet containing coherent mixtures
of two mass eigenstates with different masses and therefore  different 
momenta and/or energies. The decay interaction
conserves momentum in the laboratory system. The incident pion wave
packet must contain coherent mixtures with the same momentum difference. The
pion is a one-particle state with a definite mass. Two states with
different momenta must have different energies. A transition from a linear
combination of two states with different energies to a final state with a
single energy can occur only with a violation of energy conservation.
This violation can only occur if the $\pi$, $\mu$ and $\nu$ are not isolated
but interacting with another system that absorbs the missing energy.

A simple description of neutrino
oscillations which neglects these interactions has a missing
four-momentum expressed simply in the laboratory system as a missing energy.
Covariant descriptions and Lorentz transformations with a missing four-momentum
are not easily described in treatments which separate the decay process from
interactions with the environment.  In other Lorentz frames both energy
and momentum conservation are violated.

The momentum difference between the two coherent components of the initial
pion wave packet depends on the mass difference between neutrino mass
eigenstates. Measuring this momentum difference can give
information about the neutrino masses even if
the neutrino is not
detected. In most cases such a measurement is not
feasible experimentally. The GSI experiment\cite{gsi} describes a unique
opportunity.

 \subsection {Energy and momentum in the GSI experiment}

The search for what is known and what is hidden by quantum mechanics leads to
the energy-time uncertainty described in our simple example (\ref{timepert}). 
The short time interval between the last observation of the initial ion 
before decay
and the observed decay time enables enough violation of
energy conservation to prevent a missing mass experiment. This 
line broadening effect is demonstrated in eq.(\ref{tranprob})
and related to the line broadening of any decay observed in a time short compared
to its natural line width\cite{qm}. The decay to two final states is described
by two Breit-Wigner energy distributions  separated at long times. But in this
experiment\cite{gsi} and in  our simplified model 
(\ref{tint2}) the decay time is sufficiently short to make the
separation negligible in comparison with their broadened widths.  The
transition can occur coherently from two components of the initial state with
different energies and momenta into a same final state with a different common
energy and the same momentum difference. The sum of the transition amplitudes
from these two components of an initial state to the same final state depends
on their relative phase. Changes in this phase can produce oscillations. The
energy-time uncertainty is not covariant and defined only in the laboratory
system. Covariant descriptions and transformations from the laboratory to any
center-of-mass system are not valid for a description of neutrino oscillations.

 \subsection {Summary of what is known and hidden by quantum mechanics}

\begin{enumerate}

\item The final state has coherent pairs of states containing
neutrinos with different masses and different momenta and/or energies.

\item The initial state is a one-particle state with a
definite mass.

\item Momentum is conserved in the transition.

\item The initial state can contain coherent pairs with the same momentum
difference present in the final state state but these must have different 
energies.

\item Energy-time uncertainty hides information and prevents use of energy
conservation.

\item  The transition occurs coherently from two components of the
initial state with different
energies and momenta into a same final state with a different
common energy and the same momentum difference.

\item The relative phase between components with different energies changes
during the passage of the ion through the storage ring 
and can produce oscillations.

\end{enumerate}
A treatment of neutrino oscillations without explicit violation of
energy conservation describes a missing mass
experiment where no neutrino oscillations of any kind are allowed.

\section {The basic physics of the GSI experiment}

\subsection {A first order weak transition}

The initial state wave function  $\ket {i(t)}$ is a ``Mother" ion wave packet containing components with different momenta. Its development in time is described by an unperturbed
Hamiltonian denoted by $H_o$ which
describes the motion of the initial and final states in the electromagnetic
fields constraining their motion in a storage ring.
\beq{timedep}
\ket{i(t)} = e^{iH_ot} \ket {i(0)}
\eeq

The time $t = 0$ is defined as the time
of entry into the apparatus.
Relative phases of wave function components with different momenta are determined by localization in space at the point of entry into the apparatus. Since plane waves have equal
amplitudes over all space, these relative phases are seriously constrained by
requiring that the probability of finding the ion outside the storage ring must be zero.

A first-order weak decay is described by the Fermi Golden
Rule.
The  transition probability per unit time at time $t$ from an initial state
$\ket{i(t)}$ to a final state $\ket{f}$ is

\beq {fermi}
W(t) = {{2\pi}\over{\hbar}}|\bra{f} T \ket {i(t)}|^2\rho(E_f) =
{{2\pi}\over{\hbar}}|\bra{f} T e^{iH_ot} \ket {i(0)}|^2\rho(E_f)
\eeq
where $T$ is the transition operator and $\rho(E_f)$ is the density of final states  .
The transition operator $T$ conserves momentum.

If two components of the initial state with slightly different energies can
both decay into the same final state, their relative
phase changes linearly with time and can produce changes in the
transition matrix element.
The quantitative result and the question of whether oscillations
can be observed  depend upon the evolution of the
initial state.
The neutrino is not detected in the GSI experiment\cite{gsi}, but the information
that a particular linear combination of mass and momentum eigenstates would
be created existed in the system.
 Thus the same final state
can be created by either of three initial states that have the same momentum
difference.
Violation of energy conservation allows the decay
and provides a new method for  investigating the creation of such a coherent
state.



\subsection {Time dependence and internal clocks}

An external measurement of the time between  the initial observation and the
decay of a radioactive ion circulating in a  storage ring gives information
about the system only if an internal clock exists in the system.

\begin {enumerate}

\item An initial ion in a one-particle energy eigenstate has no
clock. Its propagation in time is completely described by a single unobservable
phase.

\item If the initial ion is in a coherent superposition of different energy
 eigenstates, the relative phase of any pair changes with energy. This phase
defines a clock which can measure the time between initial observation and
decay.

\item If the decay transition conserves energy, the final states produced by
the transition must also have different energies.

\item The decay probability  is proportional to the square of the sum of  the
transition matrix elements to all final states. There are no interference terms
between orthogonal final states with different energies and their relative phases are
unobservable.

\end{enumerate}

The probability $P_i(t)$ that the ion is still in its initial state at time $t$
and not yet decayed satisfies an easily solved differential equation,
\beq{nonexp1}
\frac {d}{dt} P_i(t) = - W(t)
 P_i(t)
; ~ ~ ~ ~ ~
\frac {d}{dt} log (P_i) = - W(t)
; ~ ~ ~ ~ ~ P_i(t) = e^{-\int W(t)dt}
\eeq

If $W(t)$ is independent of
time eq. (\ref{nonexp1}) gives an exponential decay.
The observation of a nonexponential decay implies that $W(t)$ is time dependent.
Time dependence  can
arise if the initial ion is in a coherent superposition of different energy
eigenstates, whose relative phases change with time. This phase defines a clock
which can measure the time between initial observation and decay. Since the
time $dt$ is infinitesimal, energy need not be conserved in this transition.
A non-exponential decay can occur only if there is a violation of energy
conservation. All treatments which assume energy conservation; e.g.\cite{Zoltan} will
only predict exponential decay.

$W(t)$
depends upon the unperturbed propagation of the initial state before
the time $t$ where its motion in the storage ring is
described by classical electrodynamics.
Any departure from exponential decay must come from the evolution in time of
the initial unperturbed state. This
can change the wave function at the time of the decay and therefore the value of
the transition matrix element. What happens after the decay cannot change the
wave function before the decay. Whether or not and how the final neutrino
is detected cannot change the decay rate.

\subsection {The role of Dicke superradiance}

Dicke\cite{Super} has shown that whenever two initial state components can
produce amplitudes for decay into the same final state, a linear combination
called ``superradiant" has both components interfering
constructively to enhance the transition.
The orthogonal state  called ``subradiant" has maximum destructive interference and
may even produce a cancelation.

The wave function of the initial state before the transition can contain pairs
of components with a momentum difference allowing both to decay into the
same final state.  This wave function can be expressed as a linear combination of superradiant and  subradiant states with a relative magnitude that
changes with time. The variation between superradiant and subradiant wave
functions affects the transition matrix element and can give rise to oscillations
in the decay probability. Since the momentum difference depends on the mass difference
between the two neutrino eigenstates these oscillations  can provide information about neutrino masses.

\section{Detailed analysis of a simplified model for Darmstadt Oscillations}

\subsection {The initial and final states for the transition matrix}

The initial radioactive ``Mother" ion is in a one-particle state with a
definite mass moving in a storage ring. There is no entanglement\cite{Zoltan}
since no other particles are present.  To obtain the required information
about  this initial state we need to know the evolution of the wave packet
during passage around the storage ring. This is not easily calculated. It
requires knowing the path through straight sections, bending sections and
focusing electric and magnetic fields.

The final state is  a ``Daughter" ion and a $\nu_e$ neutrino, a linear
combination of several $\nu$ mass eigenstates.  This $\nu_e$  is a complicated
wave packet containing different masses, energies and momenta. The observed
oscillations arise only from $\nu$ components with different masses and
different momenta and/or energies.

\subsection{Kinematics for a simplified two-component initial state.}

Consider the transition for
each component of the wave  packet which has a momentum $\vec P$ and energy $E$
in the initial state.  The final state has a recoil ion with momentum denoted
by $\vec P_R$ and  energy $E_R$ and a neutrino with energy $E_\nu$
and momentum  $\vec p_\nu$. If both energy and momenta are conserved,

\beq{epcons} E_R= E - E_\nu;  ~ ~ \vec P_R = \vec P - \vec p_\nu ; ~ ~
M^2 + m^2 - M_R^2 =2EE_\nu - 2\vec P\cdot\vec p_\nu
\eeq
where $M$, $M_R$ and $m$ denote respectively the masses of the mother and
daughter ions and the neutrino. We neglect transverse momenta and consider the
simplified two-component initial state for the ``mother" ion having momenta $P$
and $P + \delta P$ with energies $E$ and $E + \delta E$. The final state has
two components having neutrino momenta $p_\nu$ and $p_\nu + \delta p_\nu$ with
energies $E_\nu$ and $E_\nu + \delta E_\nu$ together with a recoil ion having
the same momentum and energy for both components. The changes in these
variables produced by a small change $\Delta (m^2)$ in the squared neutrino
mass are seen from eq. (\ref{epcons}) to satisfy the relation

\beq{delm3}
\frac{\Delta (m^2)}{2} =
 E \delta E_\nu
+ E_\nu \delta E -
P \delta p_\nu -p_\nu \delta P =
- E \delta E \cdot \left[1 - \frac {\delta E_\nu}{\delta E}+
\frac {p_\nu}{P} - \frac {E_\nu}{E}\right] \approx  - E\delta E
\eeq
where we have noted that momentum conservation in the transition requires
$P \delta p_\nu = P\delta P = E \delta E$,
$E$ and $P$ are of the order of the mass $M$ of the ion
and $p_\nu$ and $E_\nu$ are much less than $M$.
To enable coherence the two final neutrino components must have the same energy, i.e.
$\delta E_\nu = 0$. Since $\delta E \not= 0$ we are violating energy
conservation.

The relative phase $\delta \phi$ at a time t between the two states
$\ket{P}$ and $\ket{P + \delta P}$ is given by $\delta E \cdot t$.
Equation (\ref{delm3}) relates $\delta E$ to the difference
between the squared masses of the two neutrino mass eigenstates. Thus
\beq{delphipotalt}
E\cdot \delta E =-{{\Delta (m^2)}\over{2}}; ~ ~ ~ ~
\delta \phi \approx -\delta E\cdot t =
-{{\Delta (m^2)}\over{2E}}\cdot t
= - {{\Delta (m^2)}\over{2\gamma M}}\cdot t
\eeq
where $\gamma$ denotes the Lorentz factor $E/M$.

\subsection{Dicke superradiance and subradiance in the experiment}

Consider the transition from a simplified initial state for the
``mother" ion  with only two components denoted by $\ket{\vec P}$ and $\ket{\vec P +
\delta \vec P}$ having momenta $\vec P$ and $\vec P +\delta \vec P$ with energies $E$ and $E +
\delta E$. The final state denoted by
$\ket{f(E_\nu)}$ has
 a ``daughter" ion and an electron neutrino $\nu_e$
which is a linear  combination of  two
neutrino mass eigenstates denoted by $\nu_1$ and $\nu_2$  with masses $m_1$ and
$m_2$. To be coherent and produce oscillations the two components of
the final wave function must have the same neutrino
energy $E_\nu$ and
the same momentum $\vec P_R$ and  energy $E_R$ for the ``daughter" ion.
\beq{final2com}
\ket{f(E_\nu)}\equiv \ket{\vec P_R;\nu_e(E_\nu)}  =
\ket{\vec P_R;\nu_1(E_\nu)}\bra{\nu_1}\nu_e\rangle + \ket{\vec P_R;\nu_2(E_\nu)}\bra{\nu_2}\nu_e\rangle
\eeq
where  $\bra{\nu_1}\nu_e\rangle$ and $\bra{\nu_2}\nu_e\rangle$ are elements
of the neutrino mass mixing matrix, commonly expressed in terms of a
mixing angle denoted by $\theta$.
\beq{final3com}
\cos \theta \equiv \bra{\nu_1}\nu_e\rangle; ~ ~ ~
 \sin \theta \equiv \bra{\nu_2}\nu_e\rangle; ~ ~ ~\ket{f(E_\nu)}
= \cos \theta \ket{\vec P_R;\nu_1(E_\nu)}+ \sin \theta \ket{\vec P_R;\nu_2(E_\nu)}
\eeq
After a very short time two components with different initial
state energies can decay into a final state which has two components with the
same energy and a
neutrino state having two components with the same momentum difference
$\delta \vec P$ present in the initial state.

The momentum conserving transition matrix elements between the two initial
momentum components to final states with the same energy and momentum difference
$\delta \vec P$ are
\beq{transcom}
\bra{f(E_\nu)} T \ket {\vec P)} = \cos \theta \bra {\vec P_R;\nu_1(E_\nu)}T \ket {\vec P)}
;~ ~ ~
\bra{f(E_\nu)} T \ket {\vec P + \delta \vec P)} =\sin \theta \bra {\vec P_R;\nu_2(E_\nu)}T
\ket {\vec P + \delta \vec P)}
\eeq
We neglect transverse momenta and set
$\vec P\cdot\vec p_\nu \approx P p_\nu$ where $P$ and $p_\nu$ denote the
components of the momenta in the direction of the incident beam.
The  Dicke superradiance analog \cite{Super} is seen by defining superradiant and
subradiant states.
\beq{super}
\ket{Sup(E_\nu)}\equiv
\cos \theta \ket {P)} + \sin \theta \ket {P + \delta P)}; ~ ~ ~
\ket{Sub(E_\nu)}\equiv \cos \theta \ket {P + \delta P)}- \sin \theta \ket {P)}
\eeq
The transition matrix elements for these two states are then
\beq{trans}
\frac{\bra {f(E_\nu)} T \ket {Sup(E_\nu)}}{\bra{f} T \ket {P }} =[\cos \theta +
\sin \theta ]
; ~ ~ ~
\frac{\bra {f(E_\nu)} T \ket {Sub(E_\nu)}}{\bra{f} T \ket {P }} =  [\cos \theta -
\sin \theta ]
\eeq
where we have neglected the dependence of the transition operator $T$ on the
small change in the momentum $P$.
The squares of the transition matrix elements are

\beq{transsupsubsq}
\frac{|\bra {f(E_\nu)} T \ket {Sup(E_\nu)}|^2}{|\bra{f} T \ket {P }|^2} =
[1 + \sin 2 \theta ]
; ~ ~ ~
\frac{|\bra {f(E_\nu)} T \ket {Sub(E_\nu)}|^2}{ |\bra{f} T \ket {P }|^2 }=
[1 - \sin 2 \theta ]
\eeq

For maximum neutrino mass mixing, $\sin 2 \theta =1$ and
\beq{transsupsubmax}
|\bra {f(E_\nu)} T \ket {Sup(E_\nu)}|^2 =
2 |\bra{f} T \ket {P }|^2
; ~ ~ ~
|\bra {f(E_\nu)} T \ket {Sub(E_\nu)}|^2 = 0
\eeq

This is the standard Dicke superradiance in which all the transition strength
goes into the  superradiant state and there is no transition from the
subradiant state.

Thus from eq.
(\ref{super}) the initial state at time t varies periodically between the
superradiant and
subradiant states.
The period of oscillation $\delta t$ is obtained by setting  $\delta \phi \approx -2\pi$,
\beq{deltat}
\delta t \approx  {{4 \pi \gamma M}\over{\Delta (m^2)}}; ~ ~ ~
\Delta (m^2) = {{4 \pi \gamma M}\over{\delta t}} \approx
2.75 \Delta (m^2)_{exp}
\eeq
where the values $\delta t = 7 $ seconds  and
$\Delta (m^2) = 2.22\times 10^{-4} \rm{eV}^2= 2.75 \Delta (m^2)_{exp}$
are obtained from the GSI experiment and neutrino
oscillation experiments\cite{gsikienle}.

The theoretical value (\ref{deltat}) obtained with minimum
assumptions and no fudge factors is in the same ball park as the experimental
value obtained from completely different experiments. Better values obtained
from better calculations can be very useful in determining the masses and
mixing angles for neutrinos.

\subsection{Effects of spatial dependence}

The initial wave function travels through space as well as time.
In a storage ring the ion moves through straight sections, bending sections and
focusing fields.  All must be included to obtain a reliable
estimate for $\Delta (m^2)$. That this requires a detailed complicated calculation
is seen in examining two extreme cases
\begin{enumerate}

\item Circular motion in constant magnetic field.
The cyclotron frequency is independent of the momentum of the ion.
Only the time dependent term contributes to the phase and $\delta \phi^{cyc}$ is
given by eq. (\ref{delphipotalt})

\item Straight line motion with velocity $v = (P/E)\cdot t$.
The phase of the initial state at point $x$ in space and time $t$,
its change with energy and momentum changes $\delta P$ and $\delta E$
are
\beq{phisl} \phi^{SL} = P\cdot x -E\cdot t; ~ ~ ~ ~ \delta \phi^{SL} = (\delta P\cdot v -\delta E)\cdot t= \frac{P\delta P -E\delta E}{E}\cdot t=0
\eeq
\end{enumerate}

The large difference between the two results (\ref{delphipotalt} and (\ref{phisl}) indicate that a precise determination of the details of the motion of the mother ion in the storage ring is needed before
precise predictions of the squared neutrino mass difference can be made.

\subsection{A tiny energy scale}

The experimental result sets a scale  in time  of seven
seconds and a tiny energy scale
\beq{beat}
\Delta E \approx 2\pi \cdot \frac {\hbar}{7}
= 2\pi \cdot \frac {6.6\cdot 10^{-16}}{7} \approx 0.6\cdot 10^{-15}{\rm eV}
\eeq

This tiny energy difference between two waves which beat with a period of seven
seconds must be predictable from standard quantum mechanics using a scale from
another input. Another
tiny scale available in the parameters that describe this experiment is the
mass-squared difference between two neutrino mass eigenstates.
A prediction (\ref{deltat}) has been obtained from following
the propagation of the
initial state through the storage ring during the time before the decay.

That these two tiny energy scales obtained from completely different inputs are
within an order of magnitude of one another suggests some relation obtainable
by a serious quantum-mechanical calculation. We have shown here that the
simplest model relating these two tiny mass scales gives a result that differs
by only by a factor of less than three.

Many other possible mechanisms might produce oscillations. The
experimenters\cite{gsi} claim that they have investigated all of them. These
other mechanisms generally involve energy scales very different from the scale
producing a seven second period.

The observed oscillation is seen to arise from the relative phase between two
components of the initial wave function with a tiny energy difference
(\ref{beat}). These components travel through the electromagnetic fields
required to maintain a stable orbit. The effect in these fields on the relative
phase depends on the energy difference between the two components. Since the
energy difference is so tiny the effect on the phase is expected to be also
tiny and calculable.

\section{Conclusions}

Neutrino oscillations cannot occur if the momenta of all other particles
participating in the reaction are known and momentum and energy are conserved.
A complete description of the decay process must inlude the interaction with
the environent and violation of energy conservation.  A new oscillation
phenomenon providing information about neutrino mixing is obtained by following
the initial radioactive ion  before the decay. Difficulties introduced in
conventional $\nu$ experiments by tiny neutrino absorption cross sections and
very long oscillation wave lengths are avoided. Measuring the decay time
enables every $\nu$ event to be observed and counted without the necessity of
observing the $\nu$ via the tiny absorption cross section. The confinement of
the initial ion in a storage ring enables long wave lengths to be measured
within the laboratory.

\section{Acknowledgement}
The theoretical analysis in this paper was motivated by discussions with  Paul
Kienle at a very early stage of the experiment in trying to understand whether
the effect was real or just an experimental error.
It is a pleasure to thank him for calling my attention to this problem
at the Yukawa Institute for Theoretical Physics at Kyoto  University, where
this work was initiated during the YKIS2006 on ``New  Frontiers on QCD".
Discussions on possible experiments with Fritz Bosch, Walter Henning, Yuri
Litvinov and Andrei Ivanov are also gratefully acknowledged along with a
critical review of the present manuscript. The author also acknowledges further
discussions on neutrino oscillations as ``which path" experiments with  Eyal
Buks, Avraham Gal, Terry Goldman, Maury Goodman,  Yuval Grossman, Moty Heiblum, Yoseph Imry,
Boris Kayser, Lev Okun, Gilad Perez, Murray Peshkin, David Sprinzak, Ady Stern,
Leo  Stodolsky and Lincoln Wolfenstein.
%
\catcode`\@=11 
\def\references{
\ifpreprintsty \vskip 10ex
%
\hbox to\hsize{\hss \large \refname \hss }\else
\vskip 24pt \hrule width\hsize \relax \vskip 1.6cm \fi \list
{\@biblabel {\arabic {enumiv}}}
{\labelwidth \WidestRefLabelThusFar \labelsep 4pt \leftmargin \labelwidth
\advance \leftmargin \labelsep \ifdim \baselinestretch pt>1 pt
\parsep 4pt\relax \else \parsep 0pt\relax \fi \itemsep \parsep \usecounter
{enumiv}\let \p@enumiv \@empty \def \theenumiv {\arabic {enumiv}}}
\let \newblock \relax \sloppy
 \clubpenalty 4000\widowpenalty 4000 \sfcode `\.=1000\relax \ifpreprintsty
\else \small \fi}
\catcode`\@=12 
{\tighten

\end{document}